\documentclass[aps,pre,groupedaddress,showpacs]{revtex4}
\usepackage{graphicx}
\usepackage{dcolumn}
\usepackage{amssymb,textcomp}
\begin{document}

\title{ KOLKATA PAISE RESTAURANT PROBLEM IN SOME UNIFORM LEARNING STRATEGY  LIMITS }

\author{Asim Ghosh$^a$}
\email{asim.ghosh@saha.ac.in}
\author{Anindya Sundar Chakrabarti$^b$}
\email{aschakrabarti@gmail.com}
\author{Bikas K. Chakrabarti$^{a,b,c}$}
\email{ bikask.chakrabarti@saha.ac.in}
\affiliation{
$^a$Theoretical Condensed Matter Physics Division,
Saha Institute of Nuclear Physics, 1/AF Bidhannagar, Kolkata 700 064, India.
\\ $^b$Economic Research Unit, Indian Statistical Institute, 203 Barrackpore Trunk Road, Kolkata 700108, India.
\\$^c$Center for Applied
Mathematics and Computational Science
and Theoretical Condensed Matter Physics Division,
Saha Institute of Nuclear Physics, 1/AF Bidhannagar, Kolkata 700 064, India.
}

\begin{abstract}
 \noindent We study the dynamics of some uniform learning strategy limits or a probabilistic version of the  ``Kolkata Paise Restaurant'' problem, where $N$ agents choose  among $N$ equally priced but differently ranked restaurants every evening such that each agent can get dinner in the best possible ranked restaurant (each serving only one customer and the rest arriving there going without dinner that evening). We consider the learning to be  uniform among the agents and assume that each follow the same probabilistic strategy dependent on the information of the past successes in the game. The numerical results for utilization of the restaurants in some limiting cases are analytically examined.
\end{abstract}
\maketitle
\section{INTRODUCTION}
\noindent The Kolkata Paise Restaurant (KPR) problem (see \cite{kpr-09}) is a repeated game,  played between a large number of agents having no interaction among themselves. In KPR, $N$ prospective customers choose from $N$ restaurants each evening (time $t$) in a  parallel decision mode. Each restaurant have identical price but different  rank $k$ (agreed by the all the $N$ agents) and can serve only one customer. If more than one agents arrive at any restaurant on any evening, one of them is randomly chosen and is served and the rest do not get dinner that evening. Information regarding the agent distributions etc for earlier evenings  are available to everyone. Each evening, each agent makes his/her decision independent of others. Each agent has an  objective  to arrive at the highest possible ranked restaurant, avoiding the crowd so that he or she  gets dinner there. Because of fluctuations  (in avoiding herding behavior), more than one agents may choose the same restaurant and all of them, except the one randomly chosen by the restaurant, then miss dinner that evening and they are likely to change their strategy for choosing the respective restaurants next evening. As can be easily seen, no arrangement of the agent distribution among the restaurants can satisfy  everybody on any evening and the dynamics of optimal choice continues for ever. On a collective level, we look for the fraction ($f$) of customers getting dinner in any evening and also its distribution for various strategies of the game.

   It might be interesting to note here that for KPR, most of the strategies will give a low average (over evenings) value of resource utilization (average fraction $\bar f<<1$), because of the absence of mutual interaction/discussion among the agents. However, a simple (dictated) strategy, instructing each agent  go to a sequence of the ranked restaurants respectively on the first evening already and than shift by one rank  step in the in the next evenings  will automatically lead to the best optimized solution (with $ f=\bar f=1$). Also, each one gets in turn to the best ranked restaurant (with periodicity $N$). The process starts from the first evening itself. It is hard to find a strategy in KPR, where each agent decides independently (democratically) based on past experience and information, to achieve this even after long learning time.

 Let the strategy chosen by each agent in the KPR game be such that, at any time $t$, the probability $p_k(t)$ to arrive at the $k$-\rm{th} ranked restaurant is given by

\begin{equation}
\label{general}
 p_k(t)= \frac{1}{z}{\left [k^\alpha {\rm exp}{\left(-\frac{n_k(t-1)}{T}\right)}\right]}, \ \ z=\sum_{k=1}^N k^\alpha {\rm exp}{\left(-\frac{n_k(t-1)}{T}\right)},
\end{equation}
\noindent where $n_k(t-1)$ gives the number of agents arriving at the $k$-\rm{th} ranked restaurant on the previous  evening (or time $t-1$), $T$ is a noise scaling factor and $\alpha$ is an exponent. Here for $\alpha>0$ and $T>0$, the probability for any agent to choose a particular restaurant increases with its rank $k$ and decreases with the  past popularity of the same restaurant (given by the number $n_k(t-1)$ of agents arriving at that restaurant on the previous evening).   For $\alpha=0$ and $ T \rightarrow \infty$,  $p_k(t)=1/ N$ corresponds to random choice (independent of rank) case. For $\alpha=0$, $T\rightarrow 0$, the agents avoid those restaurants visited last evening and choose again randomly among the rest. For $\alpha=1$, and  $ T \rightarrow \infty$,  the game  corresponds to a strictly rank-dependent choice case. We concentrate on these three special limits.

\section{Numerical analysis}

\subsection{Random-choice }
\begin{figure}[h]
\centering
 \includegraphics[width=15 cm,bb=100 550 700 842]{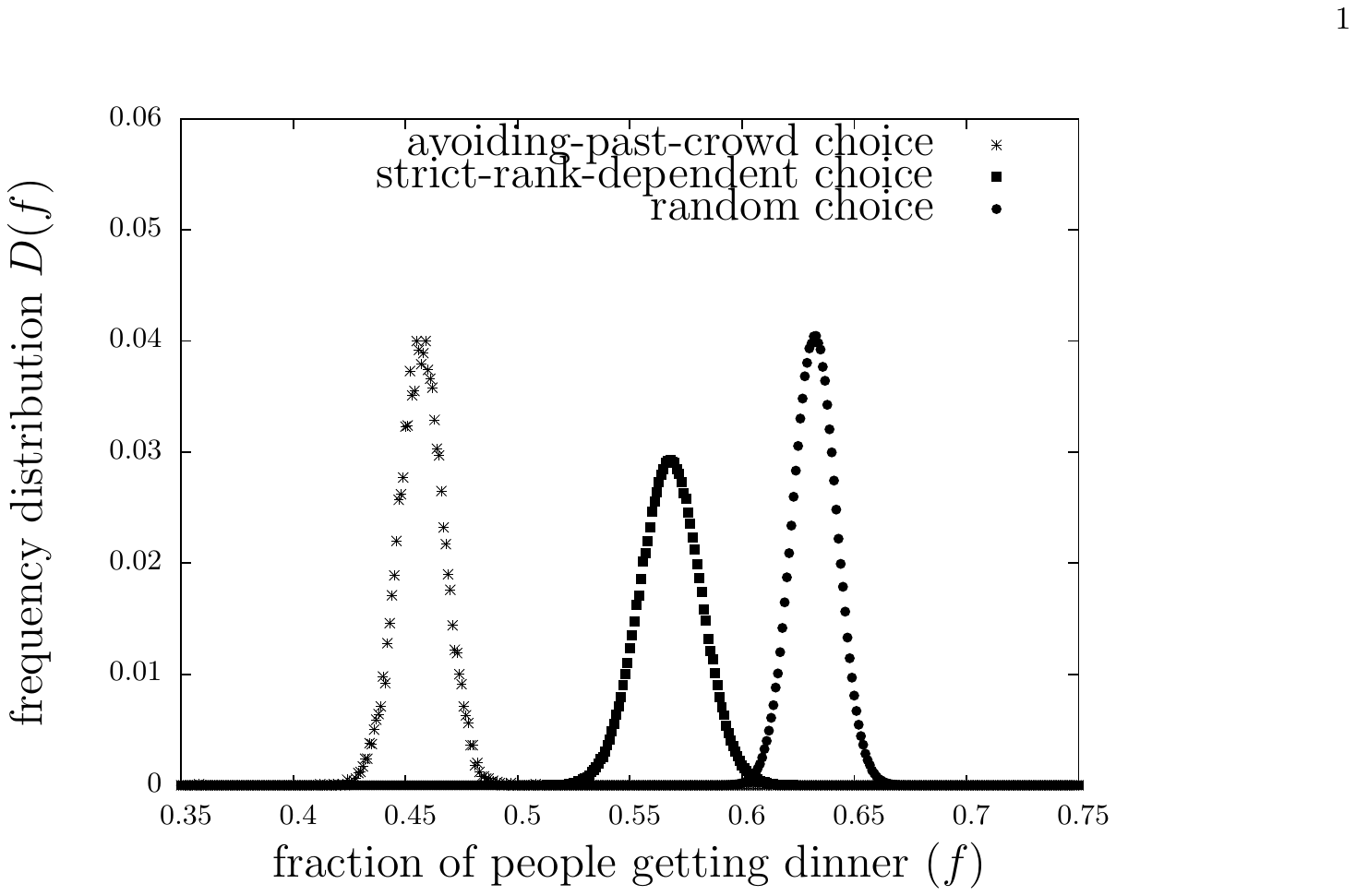}
 \caption{\label{fig:1}
 Numerical simulation results for the distribution $D(f)$ of the
 fraction $f$  of people getting dinner any evening (or fraction
 of restaurants occupied on any evening) against $f$ for different limits of $\alpha$ and $T$.
 All the  simulations have been done
 for $N=1000$ (number of restaurants and agents) and the statistics have been obtained after averages over $10^6$ time steps (evenings) after stabilization.
 }
 \end{figure}

\noindent For the case where $\alpha=0$ and $T\rightarrow\infty$, the probability $p_k(t)$ becomes independent of $k$ and becomes qeuivalent to $1/N$. For simulation  we take $1000$ restaurant and $1000$ agents and on each evening $t$ an agent selects any restaurant with equal probability $p=1/N$. All averages have been made for $10^6$ time steps. We study the variation of probability $D(f)$ of the agents getting dinner versus their fraction $f$. The numerical analysis shows that mean and mode of the  distribution occurs around $ {f} \simeq 0.63$ and that the  distribution  $D(f)$ is a Gaussian around that (see Fig. \ref {fig:1}).

\begin{figure}[h]
 \centering
 \includegraphics[width=15 cm,bb=100 550 700 842]{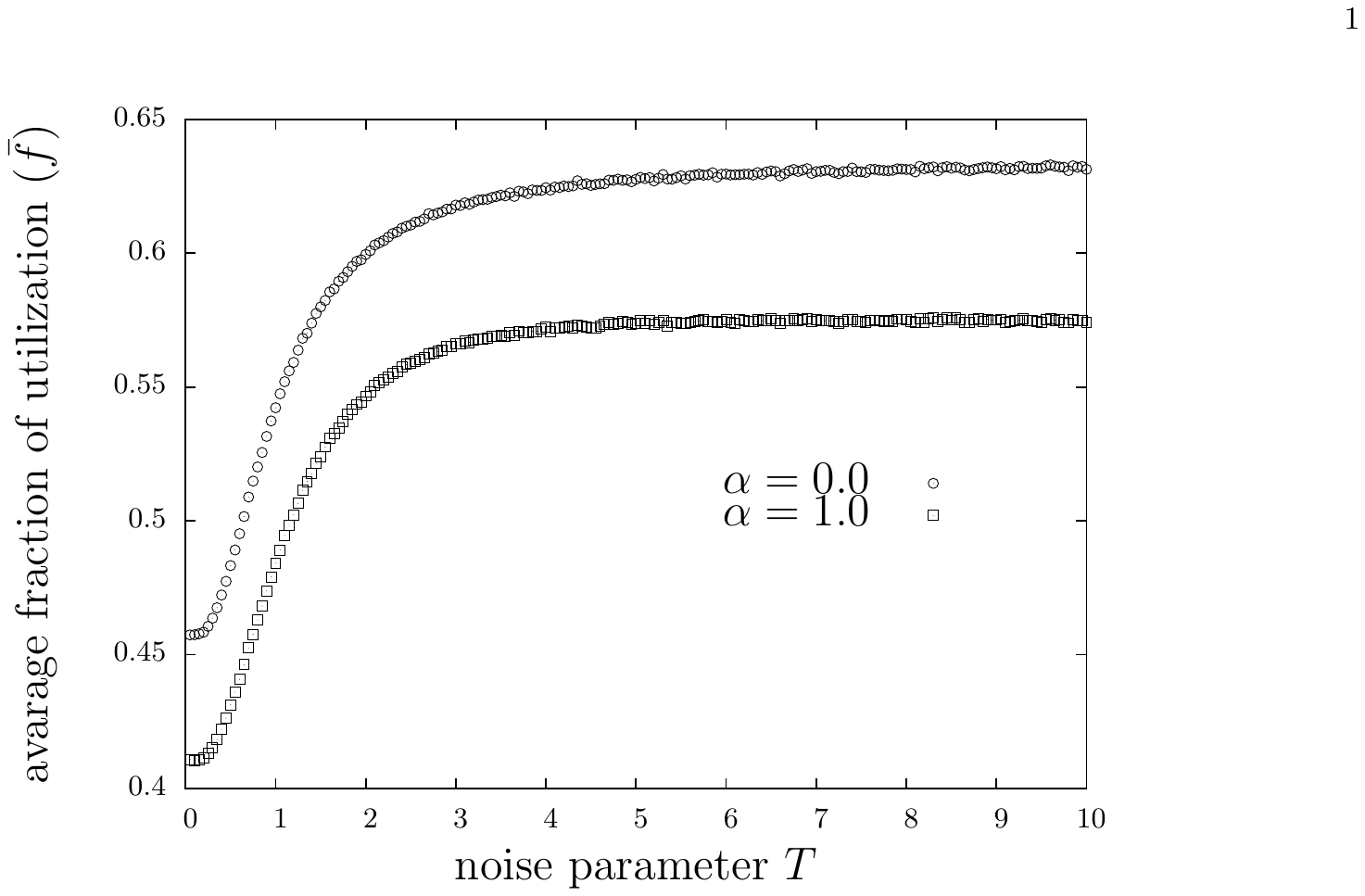}

\caption{\label{fig:2}
Numerical simulation results for the average resource utilization  fraction ($\bar f$) against the noise parameter $T$ for different values of $\alpha$ ($>$0). 
}
\end{figure}

\subsection{ Strict-rank-dependent choice}
\noindent For $\alpha=1$, $T\rightarrow\infty$, $p_k(t) = k/z$; $z=\sum k$. 
In this case, each agent chooses  a restaurant having  rank $k$ with a probability, strictly given by its rank $k$. Here also we take $1000$ agents and $1000$ restaurants and average over $10^6$ time steps for obtaining the statistics.  Fig. \ref{fig:1} shows that $D(f)$ is again a Gaussian and  that its maximum occurs at $f\simeq 0.58\equiv \bar f$.

\subsection{Avoiding-past-crowd choice}
\noindent In this case an agent chooses randomly among those restaurents which went vacant in the previous evening: with probability   $p_k(t)=  {\rm exp}(-\frac{n_k(t-1)}{T})/z$,  where  $z=\sum_k {\rm exp} (-\frac{n_k(t-1)}{T}) $ and $T\rightarrow0$,  one gets  $p_k\rightarrow0$  for $k$ values for which  $n_{k}(t-1)>0$  and $p_k=1/N'$ for other values of $k$ where $N'$ is the number of vacant restaurants in time $t-1$. For numerical studies we again take $N=1000$ and average the statistics over $10^6$ time steps. In the Fig. \ref{fig:1}, the Gaussian  distribution   $D(f)$ of restaurant utilization fraction $f$ is shown. The average utilization fraction $\bar f$ is seen to be around $0.46$. 

\section{Analytical results}

\subsection{Random-choice case}

\noindent  Suppose there are $\lambda N$ agents and $N$ restaurants. An agents can select  any restaurant with equal probability. Therefore, the probability that a single restaurant is chosen by $m$ agents is given by a Poission distribution in the limit $N\rightarrow\infty$:
\begin{eqnarray}
\label{eq:poisson_mm}
D(m) &=& \left( \begin{array}{c} \lambda N \\ m \end{array}\right)
p^m (1-p)^{\lambda N -m}; \ \ p=\frac{1}{N}  \nonumber \\
&=& \frac{\lambda^m}{m!} \exp({-\lambda}) \ \ {\rm as} \ \ N \to \infty.
\end{eqnarray}
 Therefore the fraction of restaurants not chosen by any agents is given by  $D(m=0) = \exp(-\lambda)$ and that  implies that average fraction of restaurants occupied on any evening is given by \cite{kpr-09}

 \begin{equation}
\label{rc2}
\bar{f}= 1- \exp(-\lambda) \simeq 0.63 \ {\rm for} \ \lambda=1,
\end{equation}
in the KPR problem. The distribution of the fraction of utilization will be Gaussian around this average. 

\subsection{Strict-rank-dependent choice }

\noindent In this case, an agent goes to the $k$-\rm{th} ranked restaurant with probability $p_k(t)=k/\sum k$; that is, $p_k(t)$  given by (\ref{general}) in the limit $\alpha=1$, $T\rightarrow \infty$. Starting with  $N$ restaurants and $N$ agents, we make $N/2$ pairs of restaurants and  each pair has restaurants ranked $k$ and $N+1-k$ where $1\leq k \leq N/2$. Therefore, an agent chooses any pair of restaurant with uniform probability $p=2/N$ or $N$ agents chooses randomly from $N/2$ pairs of restaurents. Therefore the fraction of pairs selected by the agents (from Eq. ~(\ref{eq:poisson_mm})) 

 \begin{equation}
f_0= 1- \exp(-\lambda) \simeq 0.86 \ {\rm for} \ \lambda=2.
\end{equation}

\noindent Also, the expected number of restaurants occupied in a pair of restaurants  with rank $k$ and $N+1-k$
 by a pair of agents is 
\begin{equation}
E_{k}=1\times\frac{k^2}{(N+1)^2}+1\times\frac{(N+1-k)^2}{(N+1)^2}+2\times2\times\frac{k(N+1-k)}{(N+1)^2}.
\end{equation}
Therefore, the fraction of restaurants  occupied by pairs of agents  
\begin{equation}
 f_1=\frac{1}{N}\sum_{i=1,...,N/2} E_{k}  \simeq 0.67.  
\end{equation}
Hence, the actual fraction of restaurants occupied by the agents is

 \begin{equation}
\bar f=f_0.f_1\simeq0.58.
\end{equation}

\noindent Again, this compares well with the numerical observation of the most probable distribution position (see Figs. \ref{fig:1} and \ref{fig:2}).

\subsection{Avoiding-past-crowd choice}
\noindent We consider here the case where  each agent chooses on any evening ($t$) randomly among the restaurants in which nobody had gone in the last evening ($t-1$). This correspond to the case where $\alpha =0$ and $T\rightarrow0$ in Eq. (\ref{general}). Our numerical simulation results for the distribution $D(f)$ of the fraction  $f$ of utilized restaurants is again Gaussian with a most probable  peak at $\bar f\simeq0.46$ (see Figs. \ref{fig:1} and \ref{fig:2}). This can be explained in the following way: As the fraction $\bar f$ of restaurants  visited by the agents in the  last evening is avoided by the agents this evening, the number of available restaurants is $N(1-\bar f)$ for this evening and is chosen randomly by all the  $N$ agents. Hence, when fitted to Eq. (\ref{eq:poisson_mm}), $\lambda=1/{(1-\bar f)}$. Therefore, following Eq. (\ref{eq:poisson_mm}), we can write the  equation for $\bar f$ as    

\begin{equation}
(1-\bar f){\left(1-{\rm exp}(-\frac{1}{1-\bar f})\right)}=\bar f .
\end{equation}

\noindent Solution of this equation gives  $\bar f\simeq0.46$. This result agrees well  with the numerical results for this limit (see Figs. \ref{fig:1} and \ref{fig:2}; $\alpha=0$, $T\rightarrow0$).

\section{Summary and discussion}

\noindent We consider here a game where $N$ agents (prospective customers) attempt to choose every  evening $(t)$ from $N$ equally priced (hence no budget consideration for any individual agent is important) restaurants (each capable of serving only one) having well-defined ranking $k$ ($=1,...,N$), agreed by all the agents. The decissions on every evening ($t$) are made by each agent independently, based on the informations about the rank $k$ of the restaurants and their past popularity given by $n_k(t-1),..,n_k(0)$ in general. We consider here cases where  each agent chooses the $k$-{\rm th} ranked restaurant with probability $p_k(t)$ given by Eq. (\ref{general}). The utilization fraction $f$ of those restaurants on every evening is studied and their distributions $D(f)$ are shown in Fig. \ref{fig:1} for some special cases. From numerical studies, we find their distributions to be Gaussian with the most probable utilization fraction $\bar f\simeq 0.63$, $0.58$ and $0.46$ for the cases with $\alpha=0$, $T\rightarrow\infty$, $\alpha=1$, $T\rightarrow\infty$ and $\alpha=0$, $T\rightarrow0$ respectively. The analytical estimates for $\bar f$ in these limits are also given and they agree very well with the numerical observations.

 The KPR problem (see also the Kolkata Restaurant Problem \cite{krp-orig}) has, in principle, a `trivial' solution (dictated from outside) where each agent gets into one of the  respective restaurant (full utilization with $f=1$) starting on the first evening  and gets the best possible sharing of their ranks as well when each one shifts to the next ranked restaurant (with the periodic boundary) in the  successive  evenings. However, this can be extremely difficult to achieve in the KPR game, even after long trial time, when each agent decides parallelly (or democratically) on their own, based on past experience  and  information regarding the history of the entire system of agents  and restaurants. The problem becomes truly difficult in the $N\rightarrow\infty$ limit. The KPR problem has similarity with the Minority Game Problem \cite{challet2005,challet2006} as in both the games, herding behavior is punished and diversity's encouraged. Also, both involves learning of the agents from the past successes etc. Of   course, KPR has some simple exact solution limits, a few of which are discussed here. In none of these cases, considered here, learning strategies are individualistic; rather all the agents choose following the probability given by Eq. (\ref{general}). In a few different limits of such a learning strategy, the average utilization fraction $\bar f$ and their distributions are obtained and compared with the analytic estimates, which are reasonably close. Needless to mention, the real challenge is to design algorithms of learning mixed strategies (e.g., from the pool discussed here) by the agents so that the simple `dictated' solution emerges eventually even when every one decides on the basis of their own information independently. 

\medskip

\noindent{\it Acknowledgment:} We are grateful to Arnab Chatterjee and Manipuspak Mitra for their important comments and suggestions.

\end{document}